# Dynamic phase transition properties for the mixed spin-(1/2, 1) Ising model in an oscillating magnetic field

**Mehmet Ertaş**[*] **and Mustafa Keskin**

*Department of Physics, Erciyes University, 38039 Kayseri, Turkey*

**Abstract**

We study the dynamic phase transition properties for the mixed spin-(1/2, 1) Ising model on a square lattice under a time-dependent magnetic field by means the effective-field theory (EFT) based on the Glauber dynamics. We present the dynamic phase diagrams in the reduced magnetic field amplitude and reduced temperature plane and find that the phase diagrams exhibit the dynamic tricritical behavior, the multicritical and zero-temperature critical points as well as reentrant behavior. We also investigate the influence of the frequency (w) and observe that for small values of w the mixed phase disappears, but high values it appears and the system displays reentrant behavior as well as critical end point.

## 1. Introduction

The mixed spin-(1/2, 1) Ising system has been studied extensively in past three decades due to the reasons that it provides a good model to investigate molecular-based magnetic materials and ferrimagnetism as well as exhibits new critical phenomena that cannot be seen in the single-spin Ising systems. The system has been used to study the equilibrium properties of different physical systems within the well-known methods in the equilibrium statistical physics [1-11] and references therein). The equilibrium critical behavior of the mixed spin-(1/2, 1) Ising (see [12-22] and references therein) and Heisenberg model [23-26] has been also investigated, extensively. The exact solution of the system was studied on different lattices, such as honeycomb lattice, a bathroom-tile or diced lattices, a Bethe lattice, two-fold Cayley tree, etc. (see [27- 36] and references therein).

On the other hand, the nonequlibrium properties of the mixed spin-1/2 and spin-1 Ising system have not been as thoroughly investigated. Godoy and Figueiredo studied the nonequilibrium behavior of the system, the Hamiltonian with only the bilinear interaction [37-39] and also including the crystal field interaction [40-41], by using the dynamical pair approximation and MC simulations. Buendía and Machado [42], and Keskin et al. [43] studied the dynamic phase transition properties and presented the dynamic phase diagrams of the mixed spin-(1/2, 1) Ising model in the presence of a time-dependent oscillating magnetic field by means the mean-field approximation (MFA) based on the Glauber-type stochastic dynamıcs. Since the spin-spin correlations are not considered in the MFA some of the first-order lines and also tricritical points in the phase diagram might be artifact of the MFA. Therefore, the dynamic phase transition properties of the system should be studied more accurate methods. Thus, the aim of this paper is to study the dynamic phase transition properties for the mixed spin-(1/2, 1) Ising model under a time-dependent magnetic field by means the effective-field theory (EFT), which considers partially spin-spin correlation, based on the Glauber-type stochastic dynamics. In particular, we study and obtain the dynamic phase transition (DPT) temperatures and present the dynamic phase diagrams.

---
[*] Corresponding author
Tel: +90 352 207 66 66#33134
E mail: mehmetertas@erciyes.edu.tr (M. Ertaş)

We should also emphasize that in recent years, the EFT based the Glauber-type stochastic dynamics have been applied to study the DPT and present the dynamic phase diagrams in the spin-1/2 [44-47], spin-1[48-50], spin-3/2 [53], spin-2 [52] and mixed spin-(2, 5/2) [53] Ising systems in detail. Moreover, in the past two decades, both experimental (see [54-58] and references therein) and theoretical (see [44-53, 59-71] and references therein) investigations of the nonequilibrium critical phenomena, especially the DPT and dynamic phase diagrams, have received a great deal of attention due to the reason that besides the scientific interests the study of DPT can also inspired new methods in materials and manufacturing process and processing as well as in nanotechnology [72].

The remainder of this article is organized as follows. In Section 2, we describe the model and its formulation. The detailed numerical results and discussions are given in Section 3. Finally Section 4 is devoted to a summary and a brief conclusion.

## 2. Model and formulation

The mixed spin-1/2 and spin-1 Ising model on a square lattice is described as a two-sublattice system, with spin variables $\sigma_i = \pm 1/2$ and $S_j = \pm 1, 0$ on the sites of sublattices A and B, respectively. The Hamiltonian of the system is given by

$$\mathcal{H} = -J \sum_{\langle ij \rangle} \sigma_i S_j - D \sum_j S_j^2 - h(t) \left[ \sum_i \sigma_i + \sum_j S_j \right], \tag{1}$$

where $\langle ij \rangle$ indicates a summation over all pairs of nearest-neighboring sites. J is the exchange interaction parameter between the two nearest-neighboring sites. D is Hamiltonian parameter and stand for the single-ion anisotropy (i.e. crystal field). h(t) is a time-dependent external oscillating magnetic field and is given by

$$h(t) = h_0 \sin(\omega t), \tag{2}$$

where $h_0$ and $\omega = 2\pi\nu$ are the amplitude and the angular frequency of the oscillating field, respectively. The system is in contact with an isothermal heat bath at absolute temperature. Within the framework of the EFT with correlations, one can easily find the magnetizations $m_A$, $m_B$ and the quadruple moment $q_B$ as coupled equations for the mixed spin-(1/2, 1) Ising system as follows

$$m_A = \langle \sigma_i \rangle = \left[ 1 + S_j \sinh(J\nabla) + S_j^2 (\cosh(J\nabla) - 1) \right]^4 F(x) \big|_{x=0}, \tag{3a}$$

$$m_B = \langle S_j \rangle = \left[ \cosh(J\nabla/2) + 2\sigma_i \sinh(J\nabla/2) \right]^4 G_1(x) \big|_{x=0}, \tag{3b}$$

$$q_B = \langle S_j^2 \rangle = \left[ \cosh(J\nabla/2) + 2\sigma_i \sinh(J\nabla/2) \right]^4 G_2(x) \big|_{x=0}, \tag{3c}$$

here $\nabla = \partial/\partial x$ is the differential operator. Hence, we obtained the set of coupled self-consistent equations. On the other hand, the self-consistent equations for m and q that are obtained by using the MFT are not coupled in the spin-(1/2, 1) model [43]. We should also mention that we have not investigated the thermal behavior of q, since we do not include the biquadratic exchange interaction parameter in Eq. (1). The functions F(x), $G_1(x)$, and $G_2(x)$ are defined as

$$F(x) = \frac{1}{2}\tanh\left[\frac{1}{2}\beta(x+h)\right], \quad (4a)$$

$$G_1(x) = \frac{2\sinh[\beta(x+h)]}{\exp(-\beta D) + 2\cosh[\beta(x+h)]}, \quad (4b)$$

$$G_2(x) = \frac{2\cosh[\beta(x+h)]}{\exp(-\beta D) + 2\cosh[\beta(x+h)]}. \quad (4c)$$

where $\beta = 1/k_B T$, $k_B$ is the Boltzman factor. Expanding the right-hand sides of Eqs. (3a)-(c), one can obtain the following equation:

$$m_A = a_0 + a_1 m_B + a_2 m_B^2 + a_3 m_B^3 + a_4 m_B^4 + a_5 m_B^5 + a_6 m_B^6 + a_7 m_B^7 + a_8 m_B^8, \quad (5)$$

$$m_B = b_0 + b_1 m_A + b_2 m_A^2 + b_3 m_A^3 + b_4 m_A^4. \quad (6)$$

In order to obtain the dynamic equations of motion for the average magnetizations, we apply the Glauber-type stochastic dynamics [73] based on the master equation as follows:

$$\frac{d}{dt}m_A = -m_A + a_0 + a_1 m_B + a_2 m_B^2 + a_3 m_B^3 + a_4 m_B^4 + a_5 m_B^5 + a_6 m_B^6 + a_7 m_B^7 + a_8 m_B^8, \quad (7)$$

and

$$\frac{d}{dt}m_B = -m_B + b_0 + b_1 m_A + b_2 m_A^2 + b_3 m_A^3 + b_4 m_A^4. \quad (8)$$

The coefficients $a_i$ (i = 0, 1, ..., 8) and $b_j$ (j = 0, 1, ..., 4) can be easily calculated employing a mathematical relation $\exp(\alpha \nabla)f(x) = f(x+\alpha)$.

### 3. Numerical results and discussions

In this section, first, we study the time dependence of average magnetizations in the mixed spin-(1/2, 1) Ising with the crystal field. The stationary solution of the dynamic equations is a periodic function of $\xi$, where $\xi = wt$, with period $2\pi$. The time dependence magnetizations $m_A(\xi)$ and $m_B(\xi)$ can be one of two types according to whether they comply with the following property or not: $m_A(\xi + 2\pi) = m_A(\xi)$ and $m_B(\xi + 2\pi) = m_B(\xi)$. At the same time, they can be one of two types according to whether they have or do not have the property

$$m_A(\xi + 2\pi) = -m_A(\xi) \text{ and } m_B(\xi + 2\pi) = -m_B(\xi). \quad (9)$$

In order to obtain the dynamic phases, we solved Eqs. (7) and (8) by using the Adams-Moulton predictor-corrector method for a given set of parameters and initial values, as presented Fig. 1. From Fig. 1(a), one can see the paramagnetic phase (p) or solutions and this solution satisfies Eq. 9. The submagnetizations $m_A$ and $m_B$ are equal to each other and oscillate around zero and are delayed with respect to the oscillating magnetic field. Fig. 1(b) shows ferrimagnetic phase (i) or solution and this solution does not satisfy Eq. 9. In this solution, the submagnetizations $m_A$ and $m_B$ are not equal to each other and $m_A$ and $m_B$ oscillate around $\pm 1/2$ and $\pm 1$, respectively. In Fig. 1(c), as $m_A$ oscillates around $\pm 1/2$, $m_B$ oscillates around $\pm 1$, which corresponds to the i phase, with the initial values of $m_A = 1/2$ and $m_B = 1$; also, $m_A$ and $m_B$ are equal to each other and they oscillate around zero, which corresponds to the p phase with the initial values of $m_A = 0.0$ and $m_B = 0.0$. Thus, we obtain coexistence solution (i + p) or the i + p mixed phase.

In order to obtain the dynamic phase boundaries among these phases and characterize the nature of the dynamic phase transitions (continuous and discontinuous), we have to investigate the thermal behavior of dynamic magnetizations ($M_{A,B}$). They are defined as

$$M_{A,B} = \frac{1}{2\pi} \int_0^{2\pi} m_{A,B}(\xi) d\xi. \qquad (10)$$

where $\xi$ represents wt. The thermal behavior of M and Q for several values of D/zJ and $h_0$/zJ are examined by combining the numerical methods of Adams-Moulton predictor corrector with the Romberg integration and their behaviors give the DPT point and the type of the dynamic phase transition. A few interesting results are given in Figs. 2(a)-(c). In these figures, $T_c$ is second order phase transition temperature and $T_t$ is a first-order phase transition temperature. In Fig. 2(a), $M_A$ and $M_B$ decrease to zero continuously as the reduced temperature (T/zJ) increases, a second-order phase transition temperature ($T_c = 0.3175$) occurs. Figs. 2(b) and 2(c) have been obtained for D/zJ = -0.375, $h_0$/zJ = 0.5375 parameters and for two different initial values, namely $M_A = 1/2$, $M_B = 1.0$ for Fig. 2(a) and $M_A = 0.0$, $M_B = 0.0$ for Fig. 2(b). In Fig. 2(b), both $M_A$ and $M_B$ undergo a first-order phase transition, because $M_A$ and $M_B$ decrease to zero discontinuously as the reduced temperature (T/zJ) increases and the phase transition is from the i phase to the p phase. Fig. 2(c) demonstrates that system does not undergo any phase transitions. Moreover, the p phase always occurs in Fig. 2(c). Therefore, the i + p mixed phase occurs below $T_t = 0.06$, in which this fact that can be seen Fig. 3(b) for $h_0$/zJ = 0.5375, explicitly.

We can now construct the dynamic phase diagrams of the model. The calculated dynamic phase diagrams are presented in the (T/zJ, $h_0$/zJ) planes for various values of the reduced crystal-field interaction (D/zJ), illustrated in Fig. 3. We also investigated the influence of longitudinal field frequency for $\omega = 0.25\pi$ and $\omega = 15\pi$, respectively and plotted one figure, namely Fig. 4. In the figures, the solid and dashed lines represent the second- and first-order phase transition lines, respectively; a filed circle denotes the dynamic tricritical point. Z, A, E, and TP are the dynamic zero-critical, multicritical, critical end, and triple points, respectively. The behavior of the phase diagrams are strongly depending on interaction parameters. As seen from Fig. 3, the following six main topological different types of phase diagrams are found and we observed four interesting phenomena. (1) The phase diagrams exhibit the p, i, and i + p phases in addition to A and Z special dynamic critical points. (2) The system displays one dynamic tricritical behavior, seen in Figs. 3(a)-(c) as well as the re-entrant behavior, seen in Fig. 3(b) and 3(e). We should also mention that several weakly frustrated ferromagnets, such as in manganite $LaSr_2Mn_2O_7$ by electron and x-ray diffraction, in the bulk bicrystals of the oxide superconductor $BaPb_{1-x}Bi_xO_3$ and $Eu_xSr_{1-x}S$ and amorphous-$Fe_{1-x}Mn_x$, demonstrate the reentrant phenomena [74-76]. (3) The dynamic tricritical behavior is not exist for large negative values of D/zJ, seen in Figs. 3(d)-(f). (4) The dynamic phase boundaries among the p and i phases are always second-order phase lines except the small values of the reduced temperature in Fig. 3(c) and the dynamic phase boundaries among the p and i + p phase are always first-order phase lines except the higher values of the reduced temperatures in Fig. 3(b).

We also studied the effect of longitudinal field frequency and presented in Figs. 4 (a) and (b) for D/zJ = -0.375, ω = 0.25π and D/zJ = -0.375, ω = 15π, respectively. If one compares Fig. 4(a) with Fig. 3(b), one can be see that i phase region becomes smaller and the dynamic tricritical point occurs for low values of $h_0$/zJ and T/zJ. Moreover, the i + p mixed phase and a special points disappear, as seen clearly in Fig. 4(a). For large values of w, the A point disappears and i + p mixed phase region also occurs for high values of T/zJ. Moreover, the system illustrates one E and TP special points as well as reentrant behavior, as seen clearly in Fig. 4(b).

## 4. Conclusions

The dynamic phase transition and dynamic phase diagrams of the kinetic mixed spin-(1/2, 1) model under a time oscillating longitudinal field are investigated using the EFT with correlations. The EFT equations of motions for the average magnetizations are obtained for the square lattice by utilizing the Glauber-type stochastic process. The dynamic phase diagrams contain the p and i fundamental phases and the i + p mixed phase as well as Z, A, E, and TP special points that strongly depend on the values of D/zJ. The system also shows dynamic tricritical and reentrant behaviors. We also find that the longitudinal field frequency (ω) greatly affects the dynamic behaviors of the system. For example, if the value of ω is high, the system shows TP and E special points instead of A special points.

Finally, in order to see the influence of the correlations, by comparing the system with the DMFT [43], the following features can be singled out: (i) While one or two dynamic tricritical points occur within the DMFT as seen in Figs. 3(a), 3(b) and Figs. 3(d)-(f) of Ref. [43], only one dynamic tricritical point exhibits within DEFT seen in Figs. 3(a)-(c). (ii) The system illustrates only Z and E special points within the DMFT but the system demonstrates Z, A, E, and TP special points. (iii) The reentrant behavior is observed by using the DEFT calculation, but not in the DMFT. (iv) The system does not undergoes a dynamic phase transition within DMFT for high negative values of D/zJ. (v) Some of the first-order phase lines either disappear or shorten within the DEFT. These facts indicate partial spin-spin correlations, which mean that thermal fluctuations, play an important role in the dynamic critical behaviors of the systems. Lastly, we hope this study will contribute to the theoretical and experimental research on the dynamic magnetic properties of kinetic mixed Ising systems as well as to research on magnetism.

**Acknowledgments**

This work was supported by Erciyes University Research Funds, Grant No. FBA-2013-4411.

**List of the figure captions**

**Fig. 1. (Color online)** Time variations of the average magnetizations ($m_A$ and $m_B$):
  a) Exhibiting a paramagnetic phase (p), $D/zJ = 0.25$, $h_0/zJ = 0.625$, and $T/zJ = 0.25$.
  b) Exhibiting a ferrimagnetic (i) phase, $D/zJ = -0.375$, $h_0//zJ = 0.375$, and $T/zJ = 0.125$.
  c) Exhibiting a mixed (i + p) phase, $D/zJ = -2.2$, $h_0//zJ = 0.175$, and $T/zJ = 0.03$.

**Fig. 2. (Color online)** The reduced temperature dependence of the dynamic magnetizations $M_A$ and $M_B$ and $T_t$ and $T_C$ are the first-order and second-order phase transition temperatures from the i phase to the p phase.

  a) Exhibiting a second-order phase transition from the i phase to the p phase for $D/zJ = -1.0$, $h/zJ = 0.1$, $\omega = 2.0\pi$; $T_c$ is found as 0.3175.

  b) **and c)** Exhibiting a first-order phase transition from the i + p phase to the p phase for $D/zJ = -0.375$, $h_0/zJ = 0.5375$; 0.06 is found $T_t$.

**Fig. 3.** Dynamic phase diagrams of the mixed spin-(1/2, 1) model in the ($T/zJ$, $h/zJ$) plane. The paramagnetic (p), ferrimagnetic (i), fundamental phases and the i + p mixed phase are obtained. Dashed and solid lines represent the first- and second-order phase transitions, respectively and the dynamic tricritical points are indicated with solid circles. Z, A, E, and TP special points are the dynamic zero temperature, multicritical, critical end point, and triple points, respectively. For $\omega = 2.0\pi$ and **a)** $D/zJ = 0.25$, **b)** $D/zJ = -0.375$, **c)** $D/zJ = -1.0$, **d)** $D/zJ = -2.0$, **e)** $D/zJ = -2.2$, and **f)** $D/zJ = -2.5$.

**Fig. 4.** Same as Fig. 3(b), but **a)** for $\omega = 0.25\pi$ and **b)** for $\omega = 25\pi$.

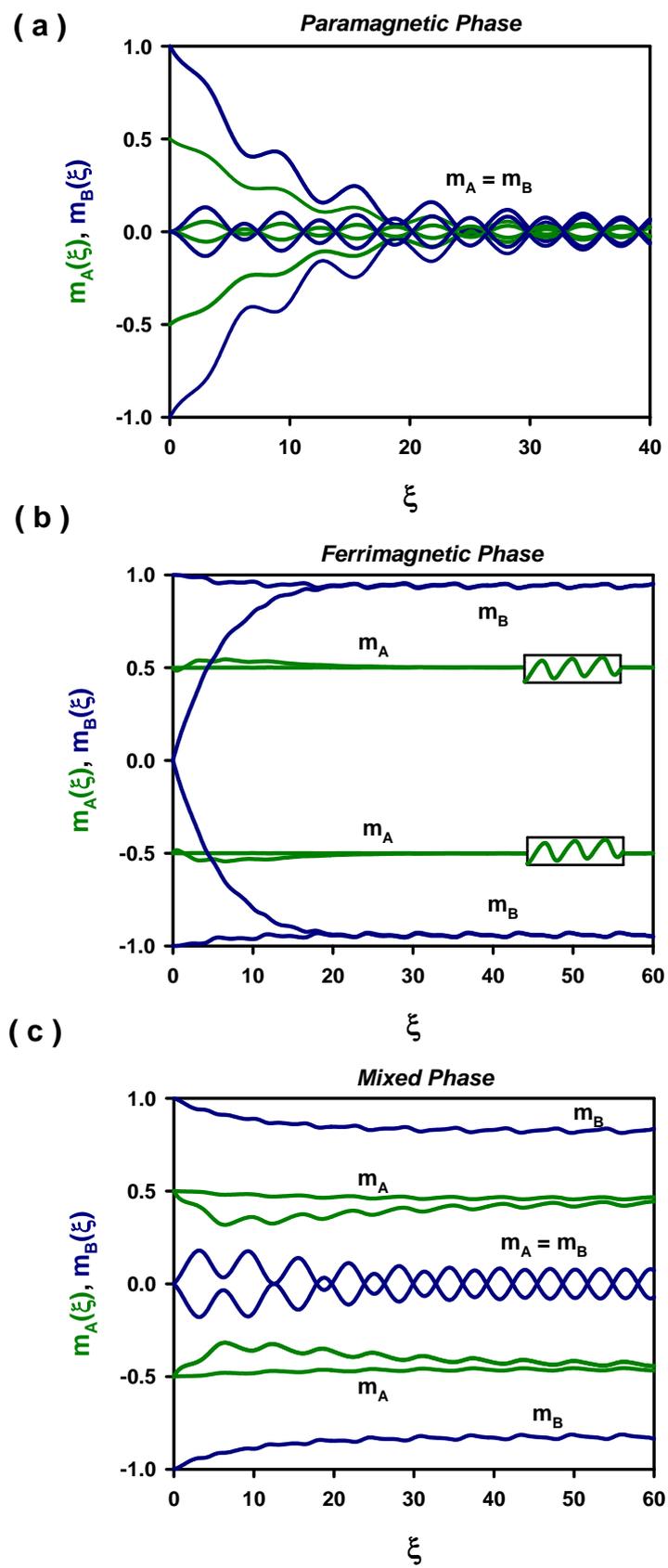

**Figure 1**

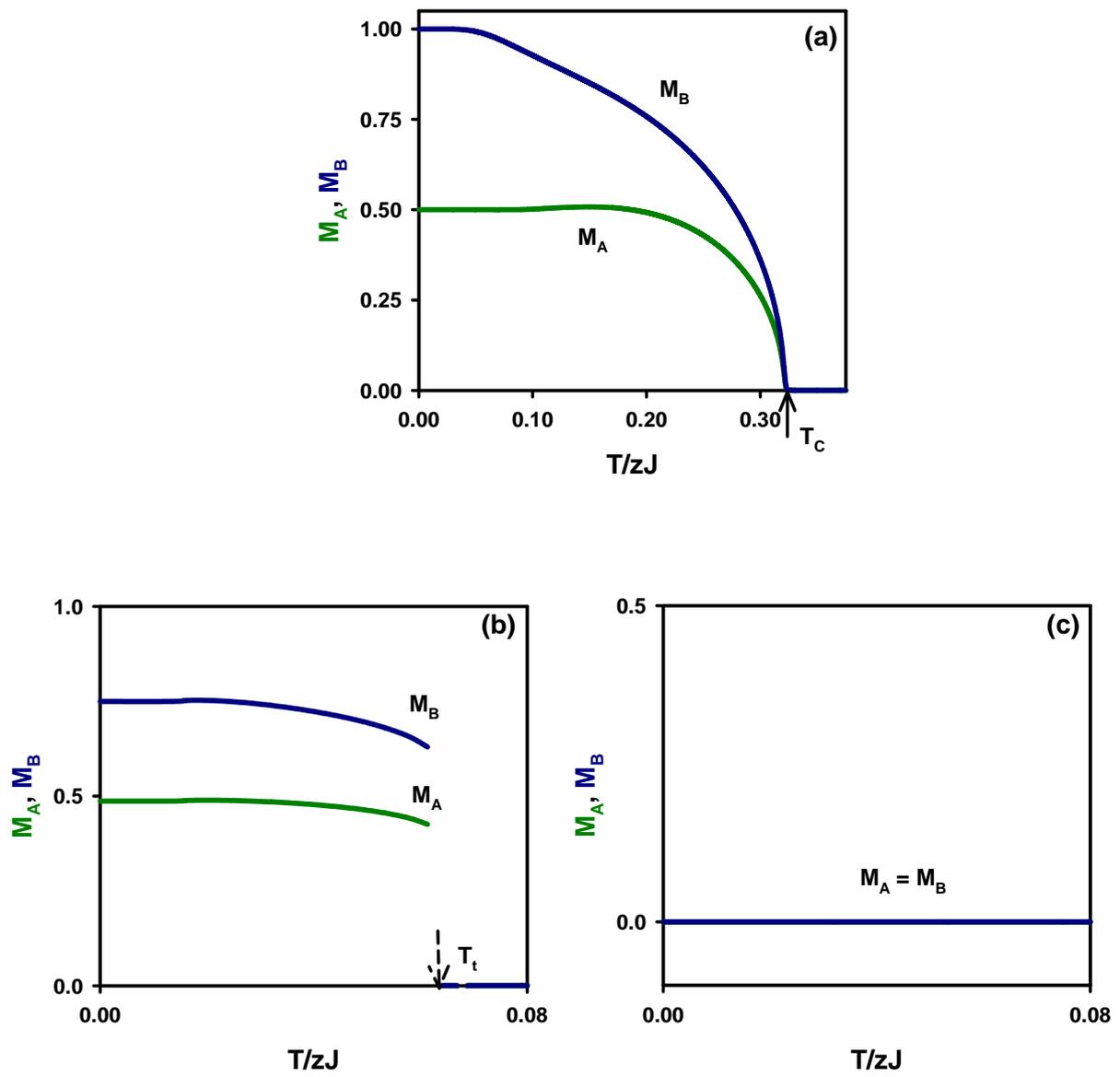

**Figure 2**

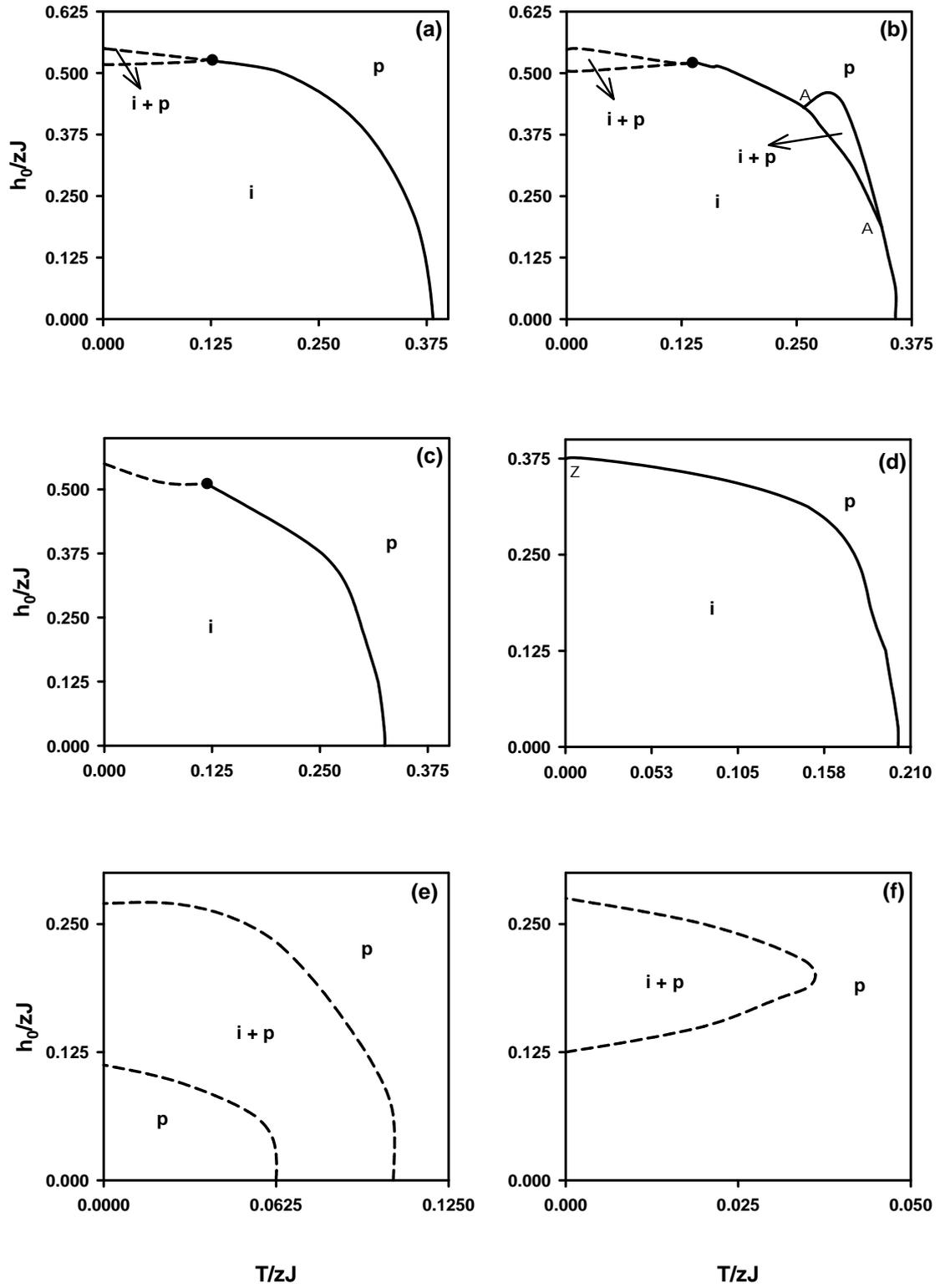

**Figure 3**

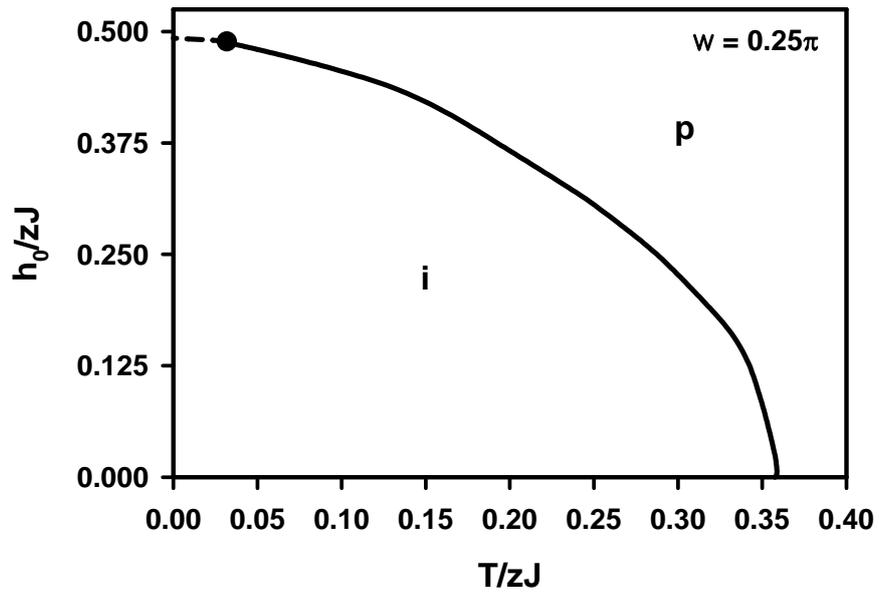

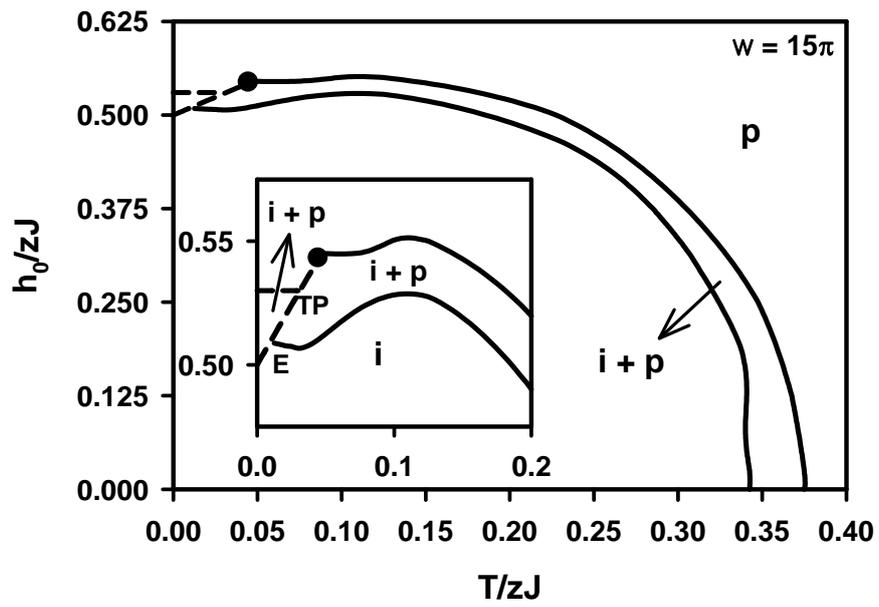

Figure 4